# "just like therapy!": Investigating the Potential of Storytelling in Online Postpartum Depression Communities

FARHAT TASNIM PROGGA, Marquette University, USA

SABIRAT RUBYA, Marquette University, USA

One in seven women encounter postpartum depression upon transitioning to motherhood. Many of them frequently seek social support in online support groups. We conducted a mixed-methods formative research to assess the potential of digital storytelling on those online platforms. We observed that mothers acquire social support from online groups through storytelling. We present design recommendations for online postpartum depression (PPD) communities to utilize storytelling in fostering emotional support and providing relevant information and education through storytelling.

**ACM Reference Format:**
Farhat Tasnim Progga and Sabirat Rubya. 2023. "just like therapy!": Investigating the Potential of Storytelling in Online Postpartum Depression Communities. In *The 2023 ACM International Conference on Supporting Group Work (GROUP '23) Companion (GROUP '23), January 8–11, 2023, Hilton Head, SC, USA.* ACM, New York, NY, USA, 4 pages. https://doi.org/10.1145/3565967.3570977

## 1 INTRODUCTION AND BACKGROUND

Women tend to use online social media to seek support during postpartum period, as social support is considered one of the vital reinforcements [3] in managing mental health. It has recently come to light that online storytelling can positively promote mutual understanding of and empathy towards lived experiences in mental health [5, 6, 8]. In this paper, we explore "online storytelling" in the context of depression (PPD) with two research questions:

**RQ1:** What are the primary themes of the stories shared on the PPD specific online communities?

**RQ2:** What benefits do women get from sharing stories on online platforms about their PPD?

To answer these questions, we used a mixed-methods strategy: quantitative and qualitative analysis of scraped data from the "Postpartum_Depression" subreddit and a qualitative analysis of semi-structured interview data.

Over the past decades, HCI, CSCW, and health informatics communities have widely analyzed content [2, 4, 5] from online mental health communities. A primary pattern observed in these mental health communities is sharing personal stories and experiences [5–8]. There is little information on if and how women with PPD use online social media platforms to tell their stories. In this formative study, we investigate the PPD storytelling behavior on a popular online platform: reddit. We find that women exchange informational and emotional support on reddit through sharing stories about a broad range of topics related to their mental health journey, and primarily seek reassurance from others. Since there is potential for digital storytelling to support PPD, we provide implications for research and design to facilitate storytelling in online communities.

| Topics | # (%) | Avg.(max) comments | Avg.(max) upvotes | Avg.(max) words in posts | Avg.(max) words in comments |
|---|---|---|---|---|---|
| Perinatal journey | **155 (15.64%)** | 6.42 **(29)** | 9.45 **(48)** | 311.8 (1050) | 133.05 (1368) |
| History of previous trauma | 127 (12.82%) | 6.00 (26) | 9.00 (33) | **448.12 (2326)** | **144.75** (953) |
| Social factors | 94 (9.49%) | 6.05 (21) | 7.62 (31) | 403.5 (1585) | 134.17 **(1666)** |
| Baby related responsibilities | 54 (5.45%) | **7.19** (18) | **9.58** (25) | 349.30 (1058) | 129.15 (561) |

Table 1. Frequency of each storytelling themes in the posts of the subreddit with average (maximum) number of comments, upvote scores, word length in the posts and comments per themes (non-exclusive).

## 2 DATA COLLECTION AND ANALYSIS

**Web scraping** We collected data from the "Postpartum_Depression" subreddit. Reddit allows users to interact with each other through upvotes, downvotes, and comments. We retrieved 1000 posts in the "new" category that spanned from March 2020 to March 2022 using the Python Reddit API Wrapper (PRAW). 992 posts and 6067 comments were considered after excluding the posts with no text or only attachment.

**Semi-structured interviews** We conducted semi-structured interviews with six PPD subreddit users to answer our RQ2. We posted our flyer on the PPD subreddit with an eligibility form. Eight mothers completed the form, and six of them were considered eligible. The interviews contained open-ended questions to capture a broad range of topic relating to their participation and online interaction. For this study, we focus on only the questions related to their experience with sharing stories and reading others' stories online (e.g., how the posts and comments from the online communities help manage their mental health issues, their experience of interaction with other community members, types of social support they look for, etc.).

**Analysis:** We conducted an inductive thematic analysis [1] of all scraped posts from the subreddit that contained aspects of storytelling. We measured the engagement of these posts through quantitative assessment of length of the posts, number and length of the comments, and upvotes ratio. We conducted deductive thematic analysis of the interview data to capture the benefits of storytelling in this particular platform.

**Ethical Considerations and Limitations:** We ensured that the posts and comments we gathered were public, and our crawling activities did not violate reddit's terms of service. The interview protocol was reviewed and approved by our university's Institutional Review Board (IRB). All participants verbally consented and allowed the interview to be audio recorded. The interview data is kept confidential, and de-identification was performed during the research.

However, the methods have inherent limitations. Analyzing only a subset of all subreddit posts across a time frame may not represent the whole PPD online storytelling dynamics. Other PPD related forums may be structured differently and may comprise different set of themes of the stories.

## 3 FINDINGS

### 3.1 Topics of the Stories in the Subreddit

Most of the 992 posts had a discussion topic and a pattern of support seeking. However, in 43.39% (n=430) of all posts, stories or experiences have been described by the poster. We found four major storytelling topics: **history of previous mental health trauma, perinatal mental health journey, social factor & baby responsibilities**. Table 1 shows the frequency of each storytelling themes in the posts of the subreddit with engagement related statistics. The themes are not mutually exclusive.

*Perinatal mental health journey* Narratives about mental health struggles were the most recurrent theme (n=155), and these posts on average gathered significant engagement, considering the maximum number of comments (29), and upvotes (48). Mothers explicitly described their struggles with mental health difficulties brought on by childbirth or pregnancy as amplifying their current mental condition, *"Ever since he was 6 months, I'm struggling with depression, I started cutting up my arms. I had this overwhelming urge to take a bunch of pills and let myself go, and I needed to do something to stop that thought because I was scared of dying. I was afraid of the pain, though, so I only cut very lightly. Then last night, it happened again, only this time, I was afraid and I made much deeper cuts."*

*History of previous trauma* We discovered that moms frequently make lengthy posts in the subreddit to express their abrasions, often connected to their delivery trauma, breastfeeding discomfort, upsetting childhood, miscarriages, or even experience of sexual assault. Such posts had moderate engagement in terms of the average number of comments (6.00) and average upvote scores (9.00).

*"Our first child died in my arms in the hospital at only 4 weeks old. After he died, I became pregnant 6 weeks later. My second pregnancy was misery. I was already depressed and deeply mourning my son... I couldn't move, sleep, BREATHE without pain due to anxiety."*

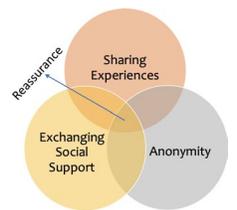

Fig. 1. Relationship among themes on the interview data

*Social factors* Moms in this online forum have highlighted how having unsupportive spouses, unhelpful families, demanding job obligations, or financial crisis has intensified their mental health concerns, *"i'm 6 weeks pp and i've had more panic attacks in the last 3 weeks than i have in the last 2 years. between my husbands god awful work schedule, my mothers false promise of help then up and getting a brand new very demanding job, and fucking exclusive pumping i am losing my mind."* These posts also generated substantial user engagement, as seen by the average number of comments on them (6.05) and the average upvotes (7.62).

*Baby related responsibilities* According to several posts, mothers' deliberate worries for their children have been illustrated. They are constantly concerned about babies' food intake, sleeping patterns, or if the child has other illnesses or unusual characteristics (e.g., autism); *"Her crying really got to me yesterday. Recently we can't even put her down in her crib she instantly stands up and won't lie down and just wails, we try to let her self soothe for 15-20 mins before getting her but I couldn't handle the crying. I never sleep and my body is \*suffering\* from sleeping with her."* The average number of comments on these posts (7.19) and the average of upvotes (9.58) show that they also drew much interaction.

### 3.2 Benefits from Storytelling

We discovered four significant themes in the interview data reflecting the moms' (n=6) experiences of storytelling: **sharing experiences, exchanging social support, anonymity, & reassurance**. Figure 1 shows the relationship of each themes in the interview data.

*Sharing experiences* Participants (n=5) reported that other community members' stories appeared more beneficial than information from other sources. They felt comfortable discussing similar situations, knowing that *"they are not alone"*. Additionally, many noted that reading other people's stories frequently helped them in ways they had not even imagined (e.g., *"just like therapy!"*).

*Exchanging social support* Women (n=6) in the interview spoke of feeling a sense of community and belonging, as well as exchanging social support as informational support *"through others experience"* or in the form of emotional support, encouragement, and praise, *"just being encouraged or having someone to talk to about my feelings is great!"*. However, it is identified by the scraped and interview data that instrumental support is less viable in such communities.

*Anonymity* Through the interview, a few participants (n=4) briefly discussed on positive and negative aspects of being anonymous (or pseudonymous) on online platforms. One of the mothers said that she occasionally has trust difficulties with pseudonyms because they seem less credible to her; she even emphasized *"I become anonymous that means I don't even trust myself."* Others, however, mentioned that the affordance of being anonymous on the platform allowed them to *"speak their mind"*.

*Reassurance* All the participants (n=6) mentioned how this community, through storytelling, has been a medium for them to seek and provide *"validation"* of their PPD thoughts and *"reassurance"* of their not being alone in this journey. Participants reported their feelings being dismissed by spouses and closed ones, while having the feeling of *"being cared for"* by sharing the same stories online with women who have gone through similar mental health experiences.

## 4 DISCUSSION AND CONCLUSION

Storytelling is prevalent in the PPD subreddit we analyzed: about half of the post we analyzed contained a story; users perceive the stories as beneficial for them. These align with other research studies [5, 6, 8] on the effectiveness of online storytelling in the context of mental health. Developers of online health communities can design the post sharing interface in a way that facilitate the storytelling process. For example, digital storytelling with use of multimedia tools can be introduced to users. Since the anonymity/pseudonimity allow users to comfortably share their experience, it is crucial that the use of multimedia tools do not conflict with the anonymity provided in these communities. Additionally, it is common in mental health communities to seek urgent support. However, there may be delay in support through comments or other interactions. If "warning signs" or "crisis events" can be detected automatically in a post, and other users do not respond promptly, the interface may display relevant stories shared previously. Storytelling can also be leveraged to educate about removing social constraints and stigma associated with PPD.